\documentclass{kapproc}
\usepackage{procps} 
\usepackage[dvips]{graphicx}
\usepackage{psfig}
\upperandlowercase
\let\footnote\savefootnote

\normallatexbib
\begin{document}
\articletitle
{Galactic magnetic fields, from radio polarimetry of the WIM}

\chaptitlerunninghead{Galactic magnetic fields}

\author{M. Haverkorn\altaffilmark{1}, P.\ Katgert\altaffilmark{2},
  A. G. de Bruyn\altaffilmark{3}, and F. Heitsch\altaffilmark{4}}

\affil{\altaffilmark{1}Harvard-Smithsonian Center for Astrophysics,
  \altaffilmark{2}Leiden Observatory, \altaffilmark{3}ASTRON/Kapteyn
  Institute, \altaffilmark{4}University of Munich}

\begin{abstract}
Multi-frequency radio polarimetry of the diffuse Galactic synchrotron
background gives new viewpoints on the Galactic magnetic
field. Rotation measure maps reveal magnetic structures on arcminute
to degree scales, such as a ring in polarization that we interpret
as a magnetic tunnel. A complication using this technique is
depolarization across the beam and along the line of sight. The
influence of beam depolarization has been estimated using numerical
models of the magneto-ionic ISM, through which polarized radiation
propagates. The models show that depolarization canals similar to
those observed can be caused by beam depolarization, and that the
one-dimensional gradients in RM needed to produce these canals are
ubiquitous in the medium.

\end{abstract}

\section{Introduction}

A proven fruitful way of probing Galactic magnetic fields is by way of
Faraday rotation in the magneto-ionic ISM. Traditionally, the
Galactic magnetic field is probed by determining the
rotation measure $RM = 0.81 \int n_e B_\parallel\, ds$ (where $n_e$ is
the thermal electron density in cm$^{-3}$, $B_{\parallel}$ is the
magnetic field component along the line of sight in $\mu$G, and $ds$
is the path length of the polarized radiation in pc) of pulsars and
linearly polarized extragalactic point sources (e.g.\ \cite{HMQ99},
\cite{CCS92}). However, due to their sparse and irregular distribution
in the sky a better background to study magnetic field structures on degree
scales with is the polarized component of the diffuse Galactic
synchrotron background.
However, depolarization of a beam (beam depolarization)
occurs if there is structure in polarization angle on scales below the
beam width. This can destroy the linear relation between polarization
angle $\phi$ and wavelength squared $\phi = RM \lambda^2$ and
therefore hamper a reliable RM determination. Furthermore, if
synchrotron emission and Faraday rotation occur in the same medium,
depolarization along the line of sight (depth depolarization) will
change the polarization characteristics as well (\cite{B66}). Depth
depolarization can also destroy the linear $\phi(\lambda^2)$ relation and
can cause apparent jumps in RM (\cite{SBS98}). Furthermore, as radiation
originating at large distances is more easily depolarized, most of the observed
polarization probes the nearby medium. The effective ``polarization
horizon'' depends on frequency and bandwidth, and varies with
position. The polarization horizon  has been estimated to be 600~pc
for 350~MHz observations (\cite{HKB03a}), and (less than) a few kpc at
1.4~GHz (\cite{LUK01}).

\section{Rotation measure rings: magnetic tunnels?}

Many observations show discrete Faraday rotation structures in
polarized intensity $P$ and/or polarization angle $\phi$ which are not
visible in total intensity, nor related to maps in any other
wavelength.  

\begin{figure}[ht]
\hbox{\psfig{figure=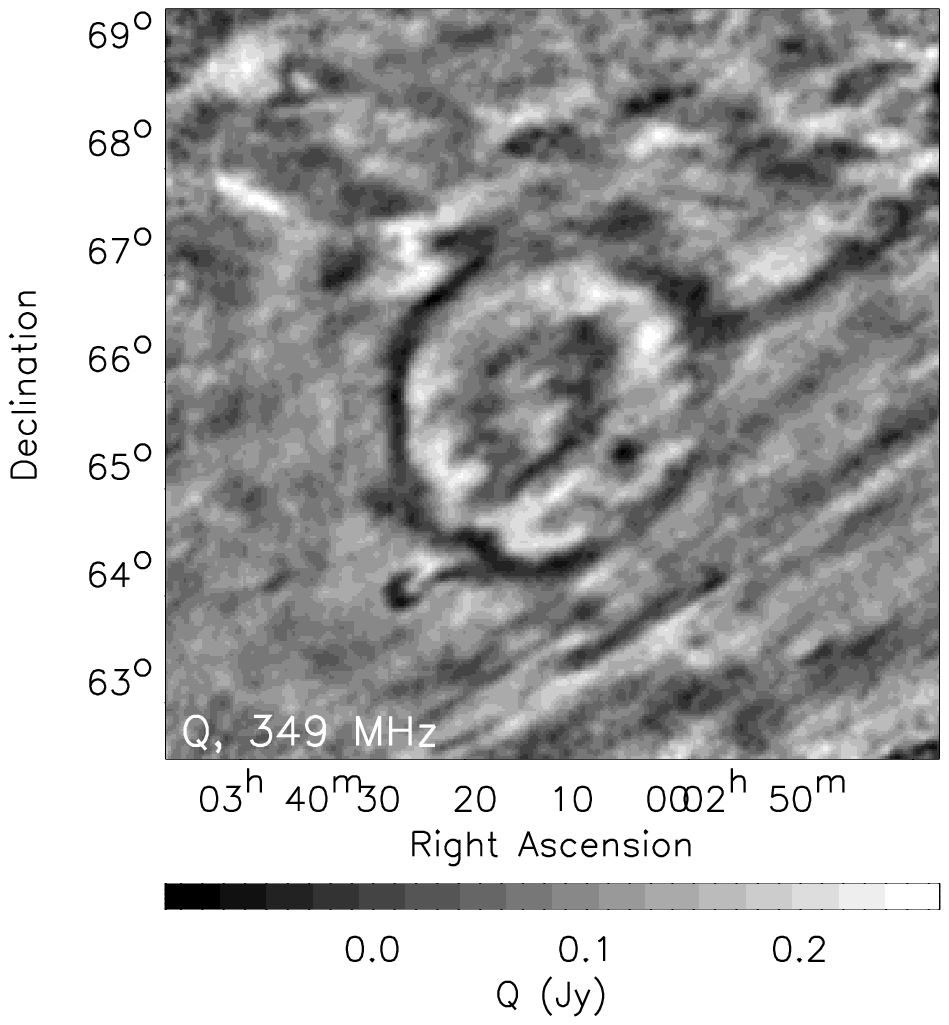,width=.25\textwidth}
      \includegraphics[width=.75\textwidth]{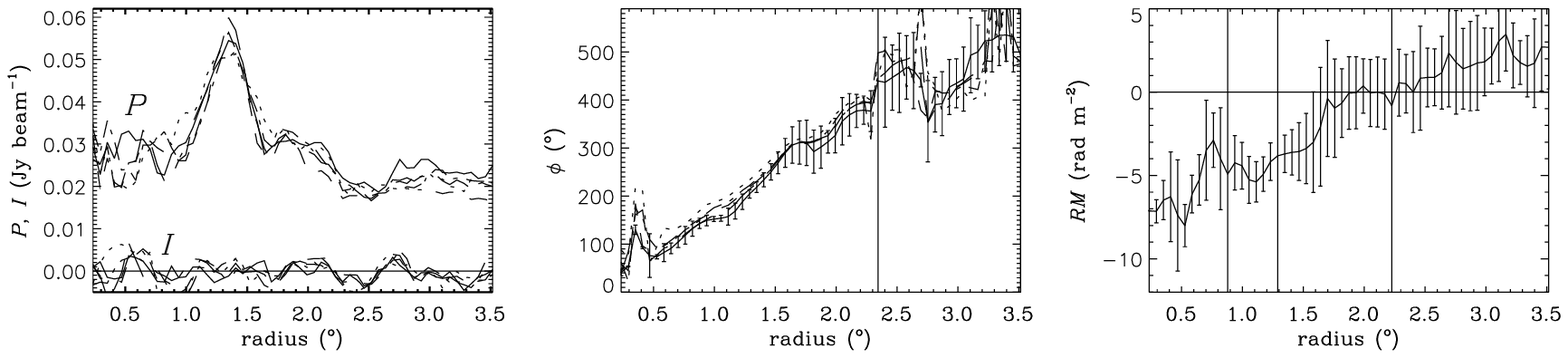}}
\caption{Left: Stokes $Q$ observed at 349~MHz with
  the Westerbork Synthesis Radio Telescope, at a resolution of $\sim
  4^\prime$. Right: Polarized intensity $P$, total intensity $I$,
  polarization angle $\phi$ and rotation measure, azimuthally averaged
  over the circular structure.}
\label{f_ringq}
\end{figure}
A particularly intriguing structure is a ring in polarization in a
$\sim 7^\circ\times7^\circ$ field centered at $(l,b) =
(137^\circ,7^\circ)$ observed with the Westerbork Synthesis Radio
Telescope (WSRT) at five frequencies around 350~MHz, discussed in
detail in \cite{HKB03b}. The structure is ring-like in $P$, and shows
a regular increase in $\phi$ from the center outwards
(Fig.~\ref{f_ringq}). This figure also shows RM, which is the most
negative in the center, increases outwards and is positive 
outside the radius of the ring in $P$. As a change in sign of RM
indicates a change in direction of $B_{\parallel}$, the direction of
the parallel magnetic field must reverse in the ring. Furthermore, the
strength of $B_{\parallel}$ and/or $n_e$ must be at a maximum
in the center of the ring. This behavior rules out any circularly
symmetric origin of the structure, such as a SNR's, stellar winds or
H{\sc II} regions, as these structures would show symmetric magnetic
field structure along the line of sight, or a maximal magnetic field
along the edges of the structure. Therefore, this structure is most
likely predominantly magnetic and funnel-like, with a maximum magnetic
field in the center directed away from the observer. As in this
orientation both $B_{\parallel}$ and $ds$ are maximal, magnetic
tunnels in other orientations may have too low RM compared to the
background to be observed. 

More circular or elliptical structures in polarization have been
observed (\cite{GLD99}, \cite{UL02}, \cite{SKH03}). As RM information is
not yet available for any of them, it is not possible to distinguish
between an electron density or magnetic field structure. If all these
structures have the same origin, magnetic tunnels of parsec size could
be fairly common in the ISM.

\section{Depolarization canals}

\begin{figure}[ht]
\includegraphics[width=.5\textwidth]{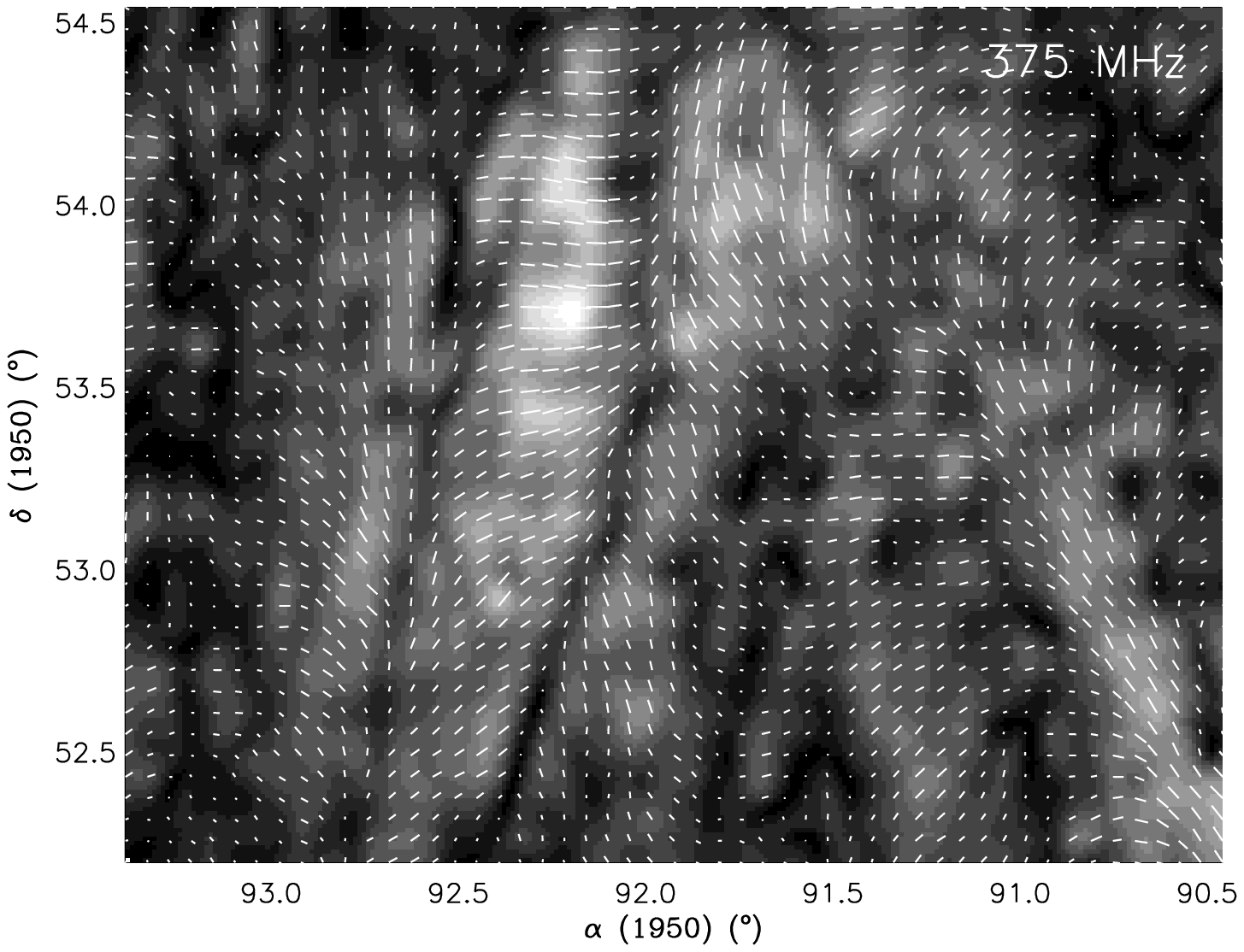}
\vspace*{-5cm}
\narrowcaption{WSRT observations of polarized intensity $P$ at 375~MHz in
  grey scale, superimposed with pseudo-vectors of polarization angle
  for independent beams (\cite{HKB03c}). The lengths of the vectors
  denote $P$. Note that the polarization angle changes by 90$^\circ$
  across depolarization canals. }
\label{f_canal}
\end{figure}

Maps of $P$ at different frequencies and resolution
show narrow extended ``depolarization canals'' (e.g.\ \cite{DHJ97},
\cite{GDM01}, \cite{UFR99}). The canals are exactly one beam wide, and
the polarization angle changes by 90$^\circ$ across them
(Fig.~\ref{f_canal}). Both beam depolarization and depth
depolarization give this signature, but the two depolarization
mechanisms demand a different underlying RM structure. If the canals
are caused by beam depolarization, RM must exhibit a sudden change of
$\Delta RM = (n+\frac{1}{2})\pi/\lambda^2$ within one beam. If depth
depolarization is responsible for the canals, the RM within a canal
must be $RM = n \pi$ \cite{SBS98}. Furthermore, canals due to beam
depolarization do not shift in position with frequency (as observed),
whereas depth depolarization canals do.

\begin{figure}[ht]
\includegraphics[width=\textwidth]{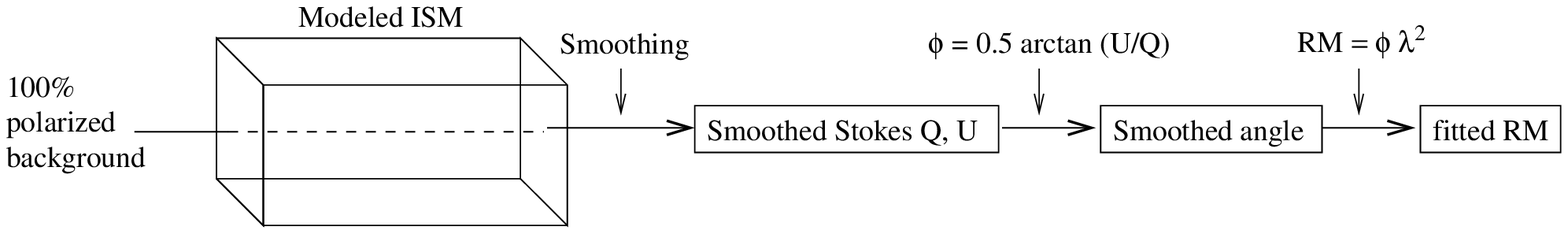}
\vspace*{-1cm}
\caption{The modeled ISM is irradiated with polarized radiation, which
  is Faraday rotated. The emerging $Q$ and $U$ distributions are
  smoothed to simulate a telescope beam, and RM is computed from the
  smoothed $Q$ and $U$ similar to the observations.}
\label{f_model}
\end{figure}

To study the true RM vs.\ the RM computed from the observations, we
constructed numerical models of the magneto-ionic ISM using the 3D Eulerian MHD
code ZEUS-3D (\cite{SN92a}, \cite{SN92b}). To
simulate a Faraday screen, the modeled ISM was irradiated with a
polarized background, see Fig.~\ref{f_model}. The ``observables'' of
the radiation after propagation through the model, Stokes $Q$ and $U$,
were smoothed with a Gaussian of width $\sigma$ pixels to simulate beam
depolarization. As there is no depth depolarization in this model, we
can estimate the effect of purely beam depolarization on the radiation
(\cite{HH03}). 

The most obvious result of the models is that they produce
depolarization canals similar to the canals observed in many $P$
maps. The canals are created at positions where the RM shows a sharp gradient in one dimension. If the magnitude of the
RM gradient is $\Delta RM =(n+\frac{1}{2})\pi/\lambda^2$ ($\approx
(n+\frac{1}{2})*4.3$~rad~m$^{-2}$ at 350~MHz), the angle
difference will be $\Delta \phi = \frac{1}{2}\pi$~rad, causing complete
depolarization. So these models show that beam depolarization can
indeed cause the observed canals. Furthermore, from the rotation measures
computed with the smoothed $Q$ and $U$ values, we conclude that for the typical
parameters of the discussed WSRT observations, beam depolarization is
expected to add an error of about 15\% to the computed RM. 

\section{Conclusions}
The RM distribution of the observed ring-like structure in polarized
intensity suggests that the structure is not spherical but a pc-scale
funnel-like structure or magnetic flux tube. Numerical models show
that sharp elongated gradients in RM, which can cause observed
depolarization canals, can exist in the warm ISM. Beam depolarization
gives an additional error of $\sim$~15\% in RM in the WSRT 350~MHz
observations discussed.

\end{document}